\documentclass[twocolumn,showpacs,preprintnumbers,amsmath,amssymb]{revtex4}

\usepackage{graphicx}
\usepackage{dcolumn}
\usepackage{bm}

\newcommand\CL{{\mathcal L}}

\newcommand\CH{{\mathcal H}}
\renewcommand\mod{~\mathrm{mod}~}

\newcommand\Nc{{Noncommutative} }
\newcommand\e{\mathrm{e}}

\begin{document}

\preprint{OIQP-06-11,  KEK-TH-1099}

\title{Noncommutative M-branes from Covariant Open Supermembranes}

\author{Makoto Sakaguchi}
 \email{makoto_sakaguchi@pref.okayama.jp}
 \affiliation{Okayama Institute for Quantum Physics, 
1-9-1 Kyoyama, Okayama 700-0015, Japan.}
\author{Kentaroh Yoshida}%
 \email{kyoshida@post.kek.jp}
\affiliation{%
Theory Division, Institute of Particle and Nuclear Studies, 
High Energy Accelerator Research
Organization (KEK), Tsukuba, Ibaraki 305-0801, Japan.
}%

\date{\today}

\begin{abstract}

We discuss an open supermembrane in the presence of a constant
three-form. The boundary conditions to ensure the $\kappa$-invariance of
the action lead to possible Dirichlet branes. It is shown that a
noncommutative (NC) M5-brane is possible as a boundary and
the self-duality condition that the flux on the world-volume satisfies
is derived from the requirement of the $\kappa$-symmetry.
We also find that the open supermembrane can attach to each of 
infinitely many M2-branes on an M5-brane,
namely a strong flux limit of the NC M5-brane.

\end{abstract}

\maketitle

Open supermembranes \cite{Strominger} are attractive
objects. Supermembranes \cite{dWHN} are considered as fundamental
objects of M-theory \cite{BFSS}, which is believed to be the unified
theory of superstrings, and open supermembranes are especially
interesting because of connection to some realistic phenomenological
models via the Horava-Witten scenario \cite{HW}. It is well-known that
an open supermembrane in flat spacetime can attach on Dirichlet
$p$-branes with $p=1,5$ and 9 \cite{EMM,dWPP}. The $p=5$ case
corresponds to M5-brane and the case with $p=9$ is nothing but the end
of the world nine-brane discussed in the Horava-Witten theory
\cite{HW}. As a recent progress, open M5-branes have also been discussed
in \cite{openM5}. 

It is interesting to consider a generalization of studies of open
supermembranes by including constant three-form gauge field (gauge field
condensate). In this direction an approach based on the light-cone
gauge, with which supermembrane theory is well studied, is not desirable
because the light-cone coordinates should inevitably be Neumann
directions and furthermore one cannot consider arbitrary electric
fluxes. Thus a covariant approach is more appropriate for 
our purpose \cite{AdS-memb}. 
In this letter we basically follow 
the covariant procedure utilized in \cite{EMM}.
This type of procedure was used to study D-branes of Green-Schwarz (GS)
string theories in \cite{LW}.

We discuss NC M5-branes 
from the requirement of the $\kappa$-symmetry of a covariant GS action
of an open supermembrane with constant fluxes. In case of open
supermembrane, the surface terms appear under the $\kappa$-variation of
the action and hence they should be deleted by imposing some appropriate
boundary conditions. The usual NC M5-branes are surely included among
the conditions, for which the self-duality condition of the flux on the
world-volume is also obtained from the condition that the projection
operator should satisfy. In addition, $p=2$ case is possible from the
argument of the boundary conditions based on the $\kappa$-symmetry.
As noted later, we should be careful for the interpretation of this
configuration. It can be supported just as a strong flux limit of NC
M5-branes and this $p=2$ case may describe infinitely many M2-branes 
on the M5-brane.

\medskip

The covariant GS action of a supermembrane in eleven-dimensional 
flat spacetime was given by \cite{BST} and it is 
composed of the Nambu-Goto (NG) part and the Wess-Zumino (WZ)
part
\begin{eqnarray}
S=\int_\Sigma d^3\xi\,\Bigl[\CL_{\rm NG} +\CL_{\rm WZ}\Bigr]\,, \nonumber 
\end{eqnarray}
where $\Sigma$ denotes the three-dimensional membrane world-volume with 
 coordinates: $\xi^i=(\xi^0=\tau,\xi^1,\xi^2)$\,. 
The NG part is given by 
\begin{eqnarray} 
\CL_{\rm NG}&=&
-\sqrt{-g(X,\theta)}~, \quad  g_{ij} = E^A_iE^B_j\eta_{AB}\,, \nonumber 
\\ 
&& E^A_i= \partial_i X^A - i\bar{\theta}\Gamma^A\partial_i\theta\,, \nonumber 
\end{eqnarray}
and the WZ part is 
\begin{eqnarray}
&& \CL_{\rm WZ}=\epsilon^{ijk}\left[
-\frac{1}{6}\partial_iX^{A} \partial_j X^{B} \partial_k X^{C} 
\CH_{ABC} \right. \nonumber \\
&& \hspace{1.2cm} +\frac{i}{2}\bar\theta\Gamma_{AB}\partial_i\theta~
 \partial_j X^A~
 \partial_kX^B
 \nonumber\\
&&   \hspace{1.2cm}
+\frac{1}{2}\bar\theta\Gamma_{AB}\partial_i\theta~
\bar\theta\Gamma^A\partial_j\theta~
\partial_kX^B \nonumber\\ 
&&   \hspace{1.2cm} \left.
-\frac{i}{6}\bar\theta\Gamma_{AB}\partial_i\theta~
\bar\theta\Gamma^A\partial_j\theta~
\bar\theta\Gamma^B\partial_k\theta \right]\,, \nonumber 
\end{eqnarray}
where $\CH \equiv C-db$ with the three-form gauge potential $C$ and 
the two-form gauge potential $b$ on the brane. 
$X^A$ ($A=0,1,\cdots,10$) denote eleven-dimensional coordinates.

\medskip 

To balance physical degrees of freedom, the action must be invariant
under the $\kappa$-variation,
\begin{eqnarray}
\delta_\kappa X^A=-i\bar\theta\Gamma^A\delta_\kappa\theta\,. \nonumber 
\end{eqnarray} 
In the open membrane case, we should be careful
with the surface terms from the variation,
while the non-surface terms vanish for a supergravity solution.
They appear only from the WZ part and so it is only for us 
to consider the $\kappa$-variation of the WZ part.

With a non-trivial constant $\CH$ along the M$p$-brane
worldvolume, the bosonic variables $X$ should satisfy 
the boundary conditions at $\partial\Sigma$ \cite{Bergshoeff} 
\begin{eqnarray}
n^i\partial_iX^{\bar{A}}
+ \epsilon^{ijk}n_i\CH^{\bar A}{}_{\bar B\bar C}\partial_j 
X^{\bar{B}}\partial_k
X^{\bar{C}}
=0\,, \nonumber 
\end{eqnarray}
for Neumann directions $\bar{A}_a~(a=0,\cdots,p)$\,, and 
\begin{eqnarray}
\partial_\tau X^{\underline{A}}=\partial_t X^{\underline{A}}=0\,, \nonumber 
\end{eqnarray}
for Dirichlet directions $\underline{A}_a (a=p+1,\cdots,10)$\,. Here
$n^i$ is a normal vector to the boundary $\partial\Sigma$, 
and the subscript $t$ implies 
a spatial direction on $\partial\Sigma$.
With the boundary conditions, the surface terms are given by 
\begin{eqnarray}
\hspace*{-0.4cm}\delta_\kappa S_{\rm WZ}|&=&\int_{\partial\Sigma} d ^2\xi
\left[
\CL^{(2)}+\CL^{(4)}+\CL^{(6)}
\right]\,,  \nonumber \\
\CL^{(2)}&=&-i
\left[
\bar\theta\Gamma_{\bar A\bar B}\delta_\kappa\theta
+\CH_{\bar A\bar B\bar C}\bar\theta\Gamma^{\bar C}\delta_\kappa\theta
\right]\dot X^{\bar{A}}{X'}^{\bar B}\,,
\label{second order in theta}
\\
\CL^{(4)}&=&
\Big[
-\frac{3}{2}\bar\theta\Gamma^{ A}\delta_\kappa\theta~
\bar\theta\Gamma_{ A\bar B}
+\frac{1}{2}\bar\theta\Gamma_{ A\bar B}\delta_\kappa\theta~
\bar\theta\Gamma^{ A}
\Big] \nonumber \\ 
&&\times (\theta'\dot X^{\bar B}-\dot \theta {X'}^{\bar B})\,, 
\label{surface term theta^4}
\\
\CL^{(6)}&=&
\frac{i}{6}
\Big[
\bar\theta\Gamma_{  A  B}\dot\theta~
\bar\theta\Gamma^{  A}\theta'~
\bar\theta\Gamma^{  B}\delta_\kappa\theta \nonumber \\ 
&& ~~~-
\bar\theta\Gamma_{  A  B}\theta'~
\bar\theta\Gamma^{  A}\dot\theta~
\bar\theta\Gamma^{  B}\delta_\kappa\theta
\nonumber\\&&~~~
-2\bar\theta\Gamma_{  A  B}\delta_\kappa\theta~
\bar\theta\Gamma^{  A}\dot\theta~
\bar\theta\Gamma^{  B}\theta'
\Big]\,, 
\label{surface term theta^6}
\end{eqnarray}
where $\dot X=\partial_\tau X$ and $X'=\partial_t X$.
$\CL^{(n)}$ denotes the terms of $n$-th order in 
$\theta$\,. 

\medskip 

To ensure the $\kappa$-invariance, (\ref{second order in theta}),
(\ref{surface term theta^4}) and (\ref{surface term theta^6}) should be
deleted by imposing some conditions on $\theta$ called ``gluing
conditions''\,. 
Let us recast surface terms to simplify our analysis, before
going to the detail.

First, with the Fierz identity 
\begin{eqnarray}
(C\Gamma_{AB})_{(\alpha\beta}(C\Gamma^A)_{\gamma\delta)}=0\,,  
\label{Fierz}
\end{eqnarray}
we rewrite $\CL^{(6)}$ as
\begin{eqnarray}
\CL^{(6)}&=&
\frac{i}{3}
\Big[
\bar\theta\Gamma_{  A  B}\dot\theta~
\bar\theta\Gamma^{  A}\theta'
-
\bar\theta\Gamma_{  A  B}\theta'~
\bar\theta\Gamma^{  A}\dot\theta~
\Big]\bar\theta\Gamma^{  B}\delta_\kappa\theta\,. \nonumber 
\end{eqnarray} 
By using (\ref{Fierz}) again
$\CL^{(6)}$ should vanish,
and thus does not affect any boundary conditions.

\medskip 

Next supposing that $\CL^{(2)}=0$\,, one can derive 
\begin{eqnarray}
 && 
\bar\theta\Gamma^{\bar A}\delta_\kappa\theta~
\bar\theta\Gamma_{\bar A\bar B}\theta'
+
\bar\theta\Gamma_{\bar A\bar B}\delta_\kappa\theta~
\bar\theta\Gamma^{\bar A}\theta' \label{cod2} \\ 
&& \hspace*{-0.3cm} =-\CH_{\bar A\bar B\bar C}
\bar\theta\Gamma^{\bar A}\delta_\kappa\theta~
\bar\theta\Gamma^{\bar C}\theta'
-\CH_{\bar A\bar B\bar C}
\bar\theta\Gamma^{\bar C}\delta_\kappa\theta~
\bar\theta\Gamma^{\bar A}\theta'
=0\,. \nonumber 
\end{eqnarray} 
Then with (\ref{Fierz}) and (\ref{cod2}) $\CL^{(4)}$ is rewritten as 
\begin{eqnarray}
&& \CL^{(4)}=-\frac{1}{2}\left[
\bar\theta\Gamma^{\underline A}\delta_\kappa\theta~
\bar\theta\Gamma_{\underline A\bar B}
+
\bar\theta\Gamma_{\underline A\bar B}\delta_\kappa\theta~
\bar\theta\Gamma^{\underline A}
\right] \times  \nonumber \\ 
&& \qquad \qquad \times (\theta'\dot X^{\bar B}-\dot \theta{X'}^{\bar B})\,.
\label{surface term theta^4 arranged}
\end{eqnarray}
Thus we have to impose such boundary conditions that delete $\CL^{(2)}$ in
(\ref{second order in theta}) and $\CL^{(4)}$ in
(\ref{surface term theta^4 arranged})\,. 
The conditions lead to possible configurations of Dirichlet branes. Let us
discuss below what configurations are allowed as consistent boundaries.

\medskip

We shall consider a Dirichlet $p$-brane by imposing a gluing
condition for the fermionic variable $\theta$ on the boundary with the
gluing matrix $M$
\begin{eqnarray}
\theta=M\theta~,~~~
M=\ell\Gamma^{\bar A_0\bar A_1\cdots\bar A_p}\,, \quad 
\ell^2(-1)^{[\frac{p+1}{2}]}s=1\,, 
\nonumber 
\end{eqnarray}
where $s=-1$ when $0\in\{\bar A_0,\bar A_1,\cdots,\bar A_p\}$ and $s=1$
otherwise. 
The matrix $M$ should satisfy $M^2=1$\,. 
Here we are considering a gluing matrix consisting of a
product of gamma matrices for the boundary condition, hence the
condition is nothing but 1/2 supersymmetric condition. We may consider,
as we will do really, various types of gluing matrices, for example, 
those composed of a sum of products of gamma matrices.

\medskip

We shall first examine $\CL^{(4)}$\,, which is independent of the
fluxes.  For $\CL^{(4)}=0$\,, 
either of the two relations 
\[
 \bar\theta\Gamma_{ \underline{A}\bar B}\delta_\kappa\theta = 0\, \quad
 \mbox{or} \quad 
 \bar\theta\Gamma^{\underline{C}}\delta_\kappa\theta =0\,, 
\] 
should be satisfied, and we can easily show that 
these imply 
\begin{eqnarray}
\text{$p=2,3\mod 4$}~~~ \mathrm{or}~~~
\text{$p=1,2\mod 4$}\,,
\label{eqn 4}
\end{eqnarray}
respectively.
Then it is turn to consider $\CL^{(2)}$\,, which depends on the
fluxes. In the case with $\CH \equiv 0$\,, $\bar\theta\Gamma_{\bar A\bar
B}\delta_\kappa\theta = 0$ is necessary for $\CL^{(2)}=0$\,. It can be
easily shown that
\begin{eqnarray}
\bar\theta\Gamma_{\bar A\bar B}\delta_\kappa\theta
=0 \quad \mbox{for}~p=1,4\mod 4\,. 
\nonumber 
\end{eqnarray}
Thus we must choose $p=1~\mod 4$\,. 

In the case with the fluxes, one might naively impose the additional
condition $\bar\theta\Gamma^{\bar C}\delta_\kappa\theta =0$\,. It can be
however shown that
\begin{eqnarray}
\bar\theta\Gamma^{\bar C}\delta_\kappa\theta =0 \quad \mbox{for}~p=3,4\mod 4
\nonumber 
\end{eqnarray}
and this condition cannot be imposed together with 
$p=1~\mod 4$\,.

\subsubsection*{M2-brane with a critical flux from NC M5-brane}

We will show that the $p=2$ case is possible under a special
circumstance, though it seems to be a bit surprising. This case is so
special that the second term in (\ref{second order in theta}) can be
rewritten as
\begin{eqnarray} 
 \CH_{\bar A_0\bar A_1\bar A_2}\bar\theta\Gamma^{\bar A_2}\delta_\kappa\theta
&=&\CH_{\bar A_0\bar A_1\bar A_2}\bar\theta\Gamma^{\bar A_2}\ell\Gamma^{\bar A_0\bar A_1\bar A_2}
\delta_\kappa\theta \nonumber \\
&=&\CH^{\bar A_0\bar A_1\bar A_2}\ell\bar\theta\Gamma_{\bar A_0\bar A_1}\delta_\kappa\theta \nonumber 
\end{eqnarray}
and thus
$\CL^{(2)}$ vanishes when
\begin{eqnarray}
1+\ell\CH^{\bar A_0\bar A_1\bar A_2}=0~.
\label{fixed H for NC M2 }
\end{eqnarray}
$\CH$ should be real so that 
$\ell$ is real
and $s=-1$.
It follows from  
(\ref{eqn 4}) 
that
$\CL^{(4)}$ 
disappears.
Thus the M2 with the critical flux $\CH$ (\ref{fixed H for NC M2
}) seems to be possible from the $\kappa$ symmetry argument. 

\medskip 

How should we interpret this configuration and the special value of the
flux? The answer is as follows. It should be considered as a strong
magnetic flux limit of a NC M5-brane. In this limit, the self-duality
condition of the flux fixes the electric flux at a critical value, and
the NC M5-brane may be seen as an M5-brane with infinitely many
M2-branes. This is an analogy to D2-brane with the infinitely large
magnetic flux, which can be seen as a D2-brane with infinitely many
D0-branes. It would be thus plausible to consider it as one of the
infinitely many M2-branes. Then the special value of the flux can be
naturally understood. In fact, if we consider the M2-brane without a
support of an M5-brane, then we confront some problems in explaining the
charge conservation law \cite{Strominger} and type IIA string
description.

\medskip 

Our analysis considers whether the boundary conditions are Neumann or
Dirichlet. In fact, three Neumann boundaries are replaced by the
Dirichlet ones in the strong flux limit. According to this, the $p=2$
case is possible in our analysis. However, our analysis does not exclude
the possibility that these exists the M5-brane behind the M2-brane.  

\medskip 

In order to confirm the above observation we next show the $p=2$ case 
is indeed realized as a strong flux limit of a NC M5-brane.

\subsubsection*{\Nc M5-brane}

We shall discuss NC M5-branes with fluxes. In comparison to a constant
NS-NS two-form in string theory, the three-form on the M5-brane
world-volume should satisfy a condition related to the self-dual
decomposition of the flux as argued by Seiberg and Witten \cite{SW}. As
we will see below, this condition appears from the $\kappa$-invariance.

\medskip 

We will consider two types of gluing matrices, (\ref{M5-1}) and
(\ref{M5-2}), as the cases (A) and (B) respectively. Both of them lead
to a commutative M5-brane in the limit the flux goes to zero. In the
strong flux limit the condition for $p=2$ is reproduced and it should
imply that the open membrane is attaching on one of the infinitely many
M2-branes on the M5-brane.

\medskip

\noindent {\bf Case (A):} 
The gluing matrix is 
\begin{align}&
M=h_0\Gamma^{\bar A_0\bar A_1\cdots\bar A_5}
+h_1\Gamma^{\bar A_0\bar A_1\bar A_2}\,.
\label{M5-1} 
\end{align}
For the condition $M^2=1$\,, which should be satisfied by the gluing
matrix, we require that
\begin{eqnarray}
&&  -s_0h_0^2-s_1h_1^2=1
\label{M5+M2: M^2=1} \\
&& s_0  = \left\{
\begin{array}{ll}
-1 ~~~~~& 0\in \{\bar A_0,\bar A_1,\cdots,\bar A_5\} \\ 
+1 & \mbox{otherwise}
\end{array}
\right.\,, \nonumber
 \\ 
&& s_1 =   \left\{
\begin{array}{ll}
-1 ~~~~~& 0\in \{\bar A_0,\bar A_1, \bar A_2 \} \\ 
+1 & \mbox{otherwise}
\end{array}
\right.\,. \nonumber
\end{eqnarray}

First, we examine $\CL^{(2)}$\,.  
After some algebra, we see that $\CL^{(2)}$ vanishes when
\begin{eqnarray}
h_1 = \CH_{\bar A_0\bar A_1\bar A_2}\,, \quad 
h_1 = h_0\CH^{\bar A_3\bar A_4\bar A_5}\,. 
\label{solution M5}
\end{eqnarray}
For reality of $\CH$, we choose $s_0=-1$.
The condition (\ref{M5+M2: M^2=1}) 
is nothing but 
the self-dual condition 
of the gauge field on the M5-brane 
\cite{SW}  
\begin{eqnarray}
\frac{1}{(\CH^{\bar A_3\bar A_4\bar A_5})^2}
-\frac{1}{(\CH_{\bar A_0\bar A_1\bar A_2})^2}
= s_1\,. \nonumber 
\end{eqnarray}
Here it should be remarked that the self-dual condition has been derived
from the $\kappa$-invariance. It would be plausible because 
the $\kappa$-symmetry of open supermembrane should be related 
to the M5-brane equation of motion \cite{CS} and 
the self-duality \cite{Sezgin}. 
For example, the solution  is
given by 
\begin{eqnarray}
\CH_{\bar A_0\bar A_1\bar A_2}= h\,, \quad 
\CH^{\bar A_3\bar A_4\bar A_5}= 
\frac{h}{\sqrt{1+s_1h^2}}\,. 
\label{sol} 
\end{eqnarray}
It is easy to see that
$\bar\theta\Gamma^{\underline A}\delta_\kappa\theta=0$ 
so that $\CL^{(4)}$ 
vanishes and
thus we have found a supersymmetric configuration
of a NC M5-brane ($\bar A_0\cdots\bar A_5$) with $\CH_{\bar
A_0\bar A_1\bar A_2}$ and $\CH^{\bar A_3\bar A_4\bar A_5}$.

\medskip 

Let us examine the two limits: 1) $\CH \to \infty$ and 2) $\CH \to
0$\,. The flux $\CH$ is assumed to be real, so it is sufficient to
examine $(s_0,s_1)=(-1,\pm 1)$\,. 

\medskip 

As the first example, we consider the case with $s_1=-1$, say, the NC M5
(012345) with $\CH_{012}$ and $\CH^{345}$. The conditions (\ref{M5+M2:
M^2=1}) and (\ref{solution M5}) are solved by
\begin{eqnarray}
&&h_0=\cos\varphi\,, \quad 
h_1=\sin\varphi \nonumber \\&&
\CH_{012}=\sin\varphi\,, \quad 
\CH^{345}=\tan\varphi \nonumber 
\end{eqnarray}
with $0\le \varphi \le\pi/2$\,, so that
\begin{eqnarray}
M=\e^{\varphi\Gamma^{345}}\Gamma^{012345}\,. \nonumber 
\end{eqnarray}
This form is very similar to the special solution found in \cite{EMM}
and it would have some relation with it. For $\varphi\to 0$\,, it
reduces to the commutative M5 (012345).  On the other hand, for
$\varphi\to\pi/2$\,, we see that $\CH^{345}\to\infty$ and so it reduces
to the previous condition for the $p=2$ case. Thus this result strongly
support our previous observation that the $p=2$ case is one of
infinitely many M2-branes on the M5-brane. This limit is nothing but the
OM limit discussed in \cite{OM} with an appropriate scaling of the
variables to make the tension finite.

\medskip

The second example is the case with $s_1=1$, the NC M5 (012345) with
$\CH_{345}$ and $\CH^{012}$\,. The conditions (\ref{M5+M2: M^2=1}) and
(\ref{solution M5}) are solved by
\begin{eqnarray}
&& h_0=\cosh\varphi\,, \quad h_1=\sinh\varphi \nonumber \\ 
&& \CH_{345}=\sinh\varphi\,, \quad \CH^{012}=\tanh\varphi \nonumber 
\end{eqnarray}
with $0\le\varphi<\infty$\,, so that
\begin{eqnarray}
M=\cosh\varphi\Gamma^{012345}+
\sinh\varphi\Gamma^{345}
=\e^{\varphi\Gamma^{012}}\Gamma^{01234 5}\,. 
\label{A-2}
\end{eqnarray}
For $\varphi\to 0$\,, it reduces to the commutative M5 (012345) since
$h_0=1$, $h_1=0$ and $\CH_{345}=\CH^{012}=0$\,. On the other hand, for
$\varphi\to\infty$\,, the boundary condition $\theta=M\theta$ is
rewritten as
\[
2\e^{-\varphi}\theta
=[(1+\e^{-2\varphi})\Gamma^{012345}+(1-\e^{-2\varphi})\Gamma^{012}
\Gamma^{012345}]\theta\,, 
\] 
which reduces to $\theta=\Gamma^{012}\theta$ in $\varphi\to\infty$\,.
Hence the previous condition for the $p=2$ case is reproduced again. 

\medskip 

\noindent {\bf Case (B):} We consider the gluing matrix
\begin{align}&
M=h_0\Gamma^{\bar A_0\bar A_1\bar A_2}
+h_1\Gamma^{\bar A_3\bar A_4\bar A_5}\,,
\label{M5-2}
\end{align}
where $0\notin \{\bar A_3,\bar A_4,\bar A_5\}$ is assumed
without loss of generality. For
$M^2=1$\,, 
\begin{eqnarray}
&& -sh_0^2-h_1^2=1\,,  \nonumber \\ 
&& s=\left\{\begin{array}{ll}
-1 ~~~~~& 0\in \{\bar A_0,\bar A_1,\bar A_2\} \\
+1 & \mbox{otherwise} 
\end{array}
\right.
\label{M2+M2: M^2=1}
\end{eqnarray}
should be satisfied. Noting the relations,
\begin{eqnarray}
\bar\theta\Gamma_{\bar A_0}\delta_\kappa\theta
=h_0\bar\theta\Gamma^{\bar A_1\bar A_2}\delta_\kappa\theta\,, \quad 
\bar\theta\Gamma_{\bar A_0 \bar A_3}\delta_\kappa\theta
=0\,, \nonumber 
\end{eqnarray}
one can easily find that $\CL^{(2)}$ vanishes when 
\begin{eqnarray}
h_0\CH^{\bar A_0\bar A_1\bar A_2}=-1~,~~~
h_1\CH^{\bar A_3\bar A_4\bar A_5}=-1~.
\label{M2+M2:solution}
\end{eqnarray}
For reality of $\CH$ we choose $s=-1$.
The condition (\ref{M2+M2: M^2=1}) with 
(\ref{M2+M2:solution}) also leads to 
the self-dual condition 
\cite{SW} 
\begin{eqnarray}
\frac{1}{(\CH^{\bar A_0\bar A_1\bar A_2})^2}
-\frac{1}{(\CH^{\bar A_3\bar A_4\bar A_5})^2}=1\,. \nonumber 
\end{eqnarray}
The solution is given by the type of (\ref{sol}) again. 
It is straightforward to see that $\CL^{(4)}$ vanishes for 
this solution 
and thus we have found a supersymmetric configuration
of a NC M5-brane. 

\medskip 

Let us consider the large/small $\CH$ limits of the brane.
For $\CH$ to be real, 
$\CH^{\bar A_0\bar A_1\bar A_2}$ has to be electric.
We consider the NC M5 (012345) with
$\CH^{012}$ and $\CH^{345}$\,. 
Equations (\ref{M2+M2: M^2=1}) and (\ref{M2+M2:solution}) are solved by 
\begin{eqnarray}
&& h_0=\cosh\varphi\,, \quad 
h_1=\sinh\varphi\,, \quad 
0\le \varphi < \infty \nonumber \\ 
&& \CH^{012}=-\frac{1}{\cosh\varphi}\,, \quad 
\CH^{345}=-\frac{1}{\sinh\varphi}\,,  \nonumber 
\end{eqnarray}
so the gluing matrix is 
\begin{eqnarray}
M=\cosh\varphi\Gamma^{012}+\sinh\varphi\Gamma^{345}
=\e^{-\varphi\Gamma^{012345}}\Gamma^{012}\,.
\label{B}
\end{eqnarray}
For $\varphi\to 0$\,, $\CH^{345}\to -\infty$ and M2 (012) with
$\CH^{012}=-1$ is left\,. On the other hand, for $\varphi\to \infty$\,,
since all the components of $\CH$ become zero and the boundary condition
reduces to $ \theta=\Gamma^{012345}\theta$\,, we obtain a commutative
M5-brane.

\medskip 

Eventually, the two cases (A) and (B), which are seemingly different, are
shown to be equivalent. The condition with (\ref{B}) can be rewritten as
\begin{eqnarray} 
&& \theta = M_d\,\theta\,, \quad M_d^2=1\,, \nonumber \\
&& M_d=\coth\varphi\Gamma^{012345}+\frac{1}{\sinh\varphi}\Gamma^{345}\,, 
\nonumber
\end{eqnarray}
so $M_d$ is another solution of (\ref{M5+M2: M^2=1}). Thus we find that
(\ref{A-2}) and (\ref{B}) describe the same NC M5-brane in different
parameterizations. This may support the possibility that the projection
operator to describe a NC M5-brane would be unique as denoted in
\cite{EMM}, though there is no proof at present. It is nice to try to
proof the uniqueness.

\medskip 

So far we have discussed NC M5-branes and M2-branes on the M5-branes as
a strong flux limit of the NC M5-branes. We can also consider
intersecting NC M5-branes from the viewpoint of $\kappa$ symmetry,
together with the idea of \cite{LW}. We will report on this issue in
\cite{soon}. It would also be interesting to study the Poisson bracket
structure in the strong flux limit along the line of
\cite{Bergshoeff,Kawamoto,Rudychev}. Here we have discussed an open
supermembrane in a strong flux limit. If we consider a closed
supermembrane instead of an open membrane in this limit, the analysis is
boiled down to that of non-relativistic limit \cite{GO} of supermembrane
\cite{GKT}. It is interesting to study the relation between
non-relativistic membrane and OM theory. It is also interesting to generalize
the study in this direction to the AdS backgrounds \cite{SY-NR}.

\begin{acknowledgements}
We would like to thank Seiji Terashima for variable comments and helpful
discussion. This work is supported in part by the Grant-in-Aid for
Scientific Research (No.~17540262 and No.~17540091) from the Ministry of
Education, Science and Culture, Japan. The work of K.~Y.\ is supported
in part by JSPS Research Fellowships for Young Scientists.
\end{acknowledgements}


\end{document}